# Spherical aberration correction in a scanning transmission electron microscope using a sculpted foil

Roy Shiloh[1]*†, Roei Remez[1]†, Peng-Han Lu[2]†, Lei Jin[2], Yossi Lereah[1], Amir H. Tavabi[2], Rafal E. Dunin-Borkowski[2], and Ady Arie[1]

[1]School of Electrical Engineering, Fleischman Faculty of Engineering, Tel Aviv University, Tel Aviv, Israel.
[2]Ernst Ruska-Centre for Microscopy and Spectroscopy with Electrons and Peter Grünberg Institute, Forschungszentrum Jülich, Jülich, Germany.
*Correspondence to: royshilo@post.tau.ac.il
†These authors contributed equally to this work.

Nearly twenty years ago, following a sixty year struggle, scientists succeeded in correcting the bane of electron lenses, spherical aberration, using electromagnetic aberration correction. However, such correctors necessitate re-engineering of the electron column, additional space, a power supply, water cooling, and other requirements. Here, we show how modern nanofabrication techniques can be used to surpass the resolution of an uncorrected scanning transmission electron microscope more simply by sculpting a foil of material into a refractive corrector that negates spherical aberration. This corrector can be fabricated at low cost using a simple process and installed on existing electron microscopes without changing their hardware, thereby providing an immediate upgrade to spatial resolution. Using our corrector, we reveal features of Si and Cu samples that cannot be resolved in the uncorrected microscope.

Electron microscopy has been revolutionised by the introduction of aberration correction, which has allowed a vast range of new scientific questions to be addressed at the atomic scale. Although the first electron microscope was built 85 years ago [1], its rotationally-symmetric, static, charge-free lenses are fundamentally incapable of correcting the dominant spherical aberration [2]. Two decades ago, Haider et al. [3,4] and Krivanek et al. [5,6] succeeded in experimentally correcting spherical aberration using electromagnetic image (objective lens) and probe (condenser lens) correctors, respectively. For the first time, researchers could obtain directly interpretable images of single atoms in both a transmission electron microscope (TEM) and a scanning TEM (STEM). With the advent of high-brightness electron guns, the improved ability to overcome noise and to resolve features in STEM images has resulted in a multitude of new applications of electron microscopy [7]. However, electromagnetic multipole aberration correctors necessitate major changes to electron column, additional space, power supplies, water cooling, and other requirements. As a result, the only practical way to improve the resolution of an uncorrected electron microscope is to replace it, at considerable cost, by a new, aberration corrected microscope. An alternative approach to improve contrast by using a phase-shifting foil was proposed as early as 1947 by Boersch [8]. Despite repeated attempts to fabricate foils of varying thickness as refractive lenses [9–11], these could not be fabricated with sufficient quality to provide viable solutions for the correction of spherical aberration. Here, 70 years later, we describe how the primary spherical aberration of the probe-forming optics in a STEM can now be corrected by using modern nanofabrication techniques to create a simple foil-based phase mask from a silicon nitride membrane [12–14], in which nm-sized variations are machined precisely using focused ion beam (FIB) milling at a tiny fraction of the complexity of an electromagnetic corrector.

Fig. 1 shows a schematic diagram of our solution, which involves passing the electron beam in a STEM through a foil of varying thickness. In an uncorrected electron microscope (Fig. 1a), the electron beam is affected by spherical aberration when it passes through the probe-forming electromagnetic lenses and limiting aperture (in our case, the Condenser 2 aperture in an FEI microscope). We replace this platinum aperture with a compact foil corrector (Fig. 1b), which introduces spherical aberration of the opposite sign, in order to form a spherical aberration corrected probe in the sample plane. To test the performance of our foil corrector, we recorded high-angle annular dark-field (HAADF) STEM images of crystalline Si in a <110> orientation both before (Fig. 1c) and after (Fig. 1d) inserting the foil corrector. The improvement in resolution is clearly visible in both real space and Fourier space, allowing 136-pm-separated dumbbells to be resolved. Fig. 1g shows the limited extent of the uncorrected spatial frequencies, in striking contrast to the corrected measurement in Fig. 1h, a result which was stable and reproducible. The images shown in Fig. 1 were post-processed using a non-linear denoising algorithm [15]. The Supplementary Material contains the raw images, as well as results recorded from an additional sample that demonstrated 128 pm resolution. Line profiles in Fig. 1e and Fig. 1f show the raw data in black and the denoised intensity in pink.

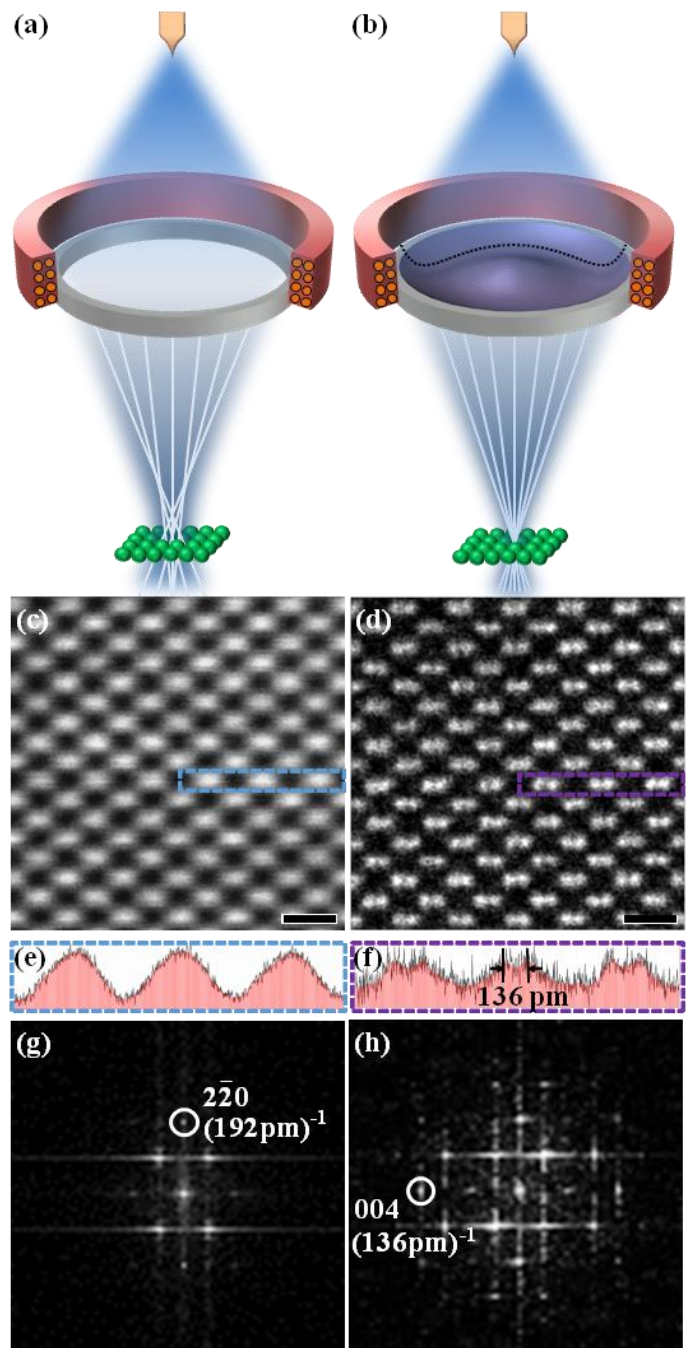

Figure 1: Schematic diagrams showing the setup for aberration correction using a foil corrector and experimental measurements. (a) An electron probe formed using a standard limiting aperture is ideally affected only by spherical aberration. (b) Equal but opposite spherical aberration is introduced by a foil corrector, yielding a spherical-aberration-free probe. The phase corrector comprises a membrane that has varying thickness along its radial direction, as indicated using a dotted black line. Without correction, Si <110> dumbbells cannot be resolved, either (c, e) in an HAADF-STEM image or (g) in its Fourier power spectrum. When using the corrector, 136 pm Si dumbbells are visible, both (d, f) in an HAADF-STEM image and (h) in its Fourier power spectrum. The images were denoised slightly (see raw data and details in the Supplementary Material). The black line profiles in (e, f) show the raw data averaged over the marked regions in (c) and (d), respectively. The scale bars in (c) and (d) are 500 pm.

Spherical aberration is the dominant aberration of electron-optical lenses. The aberrated phase can be written in the form

$$\chi = \frac{\pi}{2\lambda} C_s(\theta^4 - NA^2\theta^2),  \quad\quad 1$$

where $\theta$ is the semi-angle of the beam with respect to the optical axis, $NA = \sin\alpha$ is the numerical aperture, $\alpha$ is the convergence semi-angle, $C_S$ is the spherical aberration coefficient, and $\lambda$ is the electron wavelength. We define a $2\pi$ phase shift as *one cycle*, abbreviated "$1\lambda$", and characterize the maximal phase difference introduced by the aberrated phase $\chi$ using the figure of merit "*peak-to-valley*" [16], which is calculated to be $\Lambda = C_s NA^4/16\lambda$ cycles. For a non-magnetic material, the electron beam accumulates a phase shift that depends on the foil thickness $t$ according to the expression

$$\varphi = C_E U_i t, \quad\quad 2$$

where $U_i$ is the mean inner potential of the foil and $C_E$ is a constant that depends on the electron energy. In order to negate spherical aberration, the thickness profile of the foil must satisfy the relation $\varphi = -\chi$. If $t_{2\pi}$ is defined to be the required thickness for a phase shift of one cycle, then the thickness profile of the foil corrector is given by the expression

$$t(r) = t_{2\pi} \times \frac{4\Lambda}{r_{max}^4}(r^4 - r^2 r_{max}^2), \quad\quad 3$$

where $r_{max}$ is the radius of the phase corrector's limiting aperture, or aperture stop, i.e., the radius that defines the numerical aperture of the probe-forming optics. A small fraction of the foil thickness supports the structure mechanically and remains unprocessed. We fabricated our spherical aberration foil corrector using FIB milling and installed it in the Condenser 2 aperture position in a 300 kV ($\lambda$ = 1.97 pm) FEI Titan microscope [17], in which the uncorrected spherical aberration coefficient $C_s$ of the probe-forming lenses is known to be 2.7 mm. This microscope is equipped with a CEOS post-specimen spherical aberration corrector. However, it cannot correct the pre-specimen illumination optics (the probe-forming optics) and therefore has no advantageous effect on STEM measurements.

The spherical aberration in a STEM probe can be inferred from an electron Ronchigram [18], which involves recording the diffraction pattern of a stationary probe that is focused on the sample, in our case an amorphous C film. Analysis of this image reveals information about the numerical aperture of the probe and its aberrations - in particular, spherical aberration [19], which can be recognized by radii and azimuths of infinite magnification. The latter feature is evident in a Ronchigram recorded using our uncorrected microscope (Fig. 2a), which corresponds to the HAADF-STEM measurement shown in Fig. 1c. A focused aberration-free probe produces a Ronchigram that takes the form of a flat, nearly-uniformly-illuminated disc and is associated with a nearly-constant phase distribution, similar to the image shown in Fig. 2b, which was recorded using our foil corrector. According to a calculation by Scherzer [20], the numerical aperture for optimal resolution is $1.41(\lambda/C_s)^{1/4} \approx 7.34$ mrad, which is close to the value of 7.46 mrad used in our experiment. Our corrector, in its present form, is designed to correct spherical aberration at and up to a numerical aperture of ≈12 mrad, as seen in Fig. 2b. The Supplementary Material contains further details, as well as a focal series of corrected Ronchigrams. Fig. 2 also shows direct measurements of images of uncorrected (Fig. 2c) and corrected (Fig. 2d) electron probes and their profiles (Fig. 2g), which show a decrease of more than 30% in probe diameter. In order to image such a probe, the microscope should be aberration-free between the sample plane and the detector. In the present study, the post-specimen CEOS corrector can correct monochromatic image aberrations up to $\approx 15\ mrad$. Therefore, our direct probe measurements are expected to be only slightly smeared due to chromatic aberration of the objective lens. We also used the measured probes to simulate STEM images of two points that are separated by 136 pm, representing Si dumbbells, confirming the formation of irresolvable and resolvable images in Fig. 2e and Fig. 2f, respectively.

One of the primary challenges in this concept is that of scattering by the material-based corrector, which is manifested in the Ronchigram in Fig. 2b as a faint amplitude modulation, a result of the annulus thickness variation of the corrector (see also Fig. S1a). This is the only mechanism that affects otherwise 100%-efficient intensity transfer to the corrected probe. We used electron energy-loss spectroscopy (EELS) to determine [21] the inelastic mean free path in silicon nitride at 300 kV to be ≈170 nm, with an energy distribution exhibiting typical losses of 10 eV-50 eV. The angular distribution of inelastic scattering leaving the corrector has an axially sharp peak and a wide tail that reaches far beyond 50 mrad, resulting in a noisy background and a loss of contrast in the STEM image. Whereas the intensity in the sharp peak is only appreciable for angles $\theta$ that are much smaller than the characteristic angle for elastic scattering [22], the spatial distribution in the sample plane in proximity to the probe is negligible, since the typical transverse distance of scattered electrons $\theta f$ is several orders of magnitude larger than the diameter of the probe, where $f$ is the equivalent probe-forming optics' focal length. Since most contrast degradation is due to the large-area inelastic tail, it could in principle be cut off by mounting a small aperture on the sample itself, or by designing the probe-forming optics with another aperture at the position of a crossover before the sample. The inelastic scattering could also be mitigated and the STEM image contrast improved by using a thinner or fractured (modulo-$2\pi$) design [13], as discussed in the Supplementary Material.

Foil-based correctors of the present design are advantageous in that fabrication is independent of the intended accelerating voltage and that there is no strict tolerance on the profile thickness. A change in these parameters will only result in a shift of the operating numerical aperture of the corrector and the strength of correction. The corrector will then negate the spherical aberration for the new numerical aperture value. A further advantage of our corrector is that it corrects the electron probe on-axis, requiring practically no change in the standard measurement procedure.

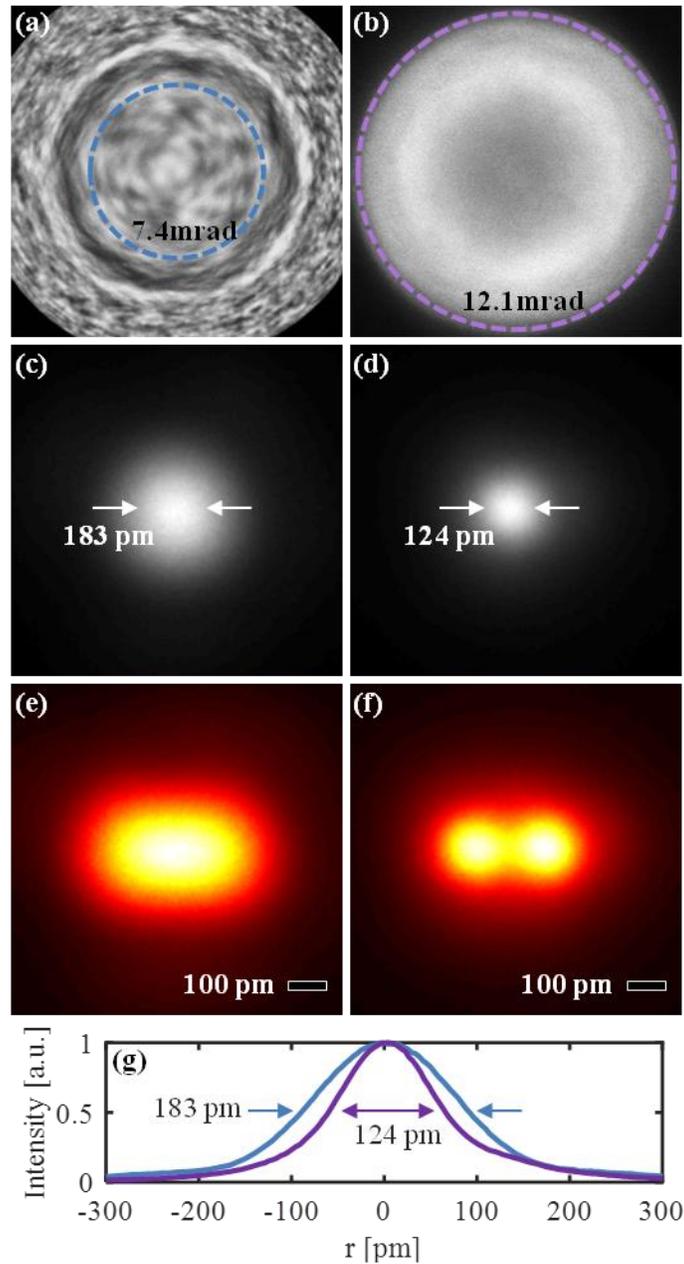

Figure 2: Assessment of correction. (a, b) Ronchigrams recorded from amorphous C for (a) the uncorrected microscope limited by spherical aberration to ≈7.4 mrad and (b) the corrected microscope for 12.1 mrad. (c, d) Full-width at half-maximum measurements of (c) an uncorrected probe for 7.46 mrad (183 pm) and (d) a corrected probe for 12.1 mrad (124 pm). (e, f) Simulation of 136-pm-separated points for (e) the uncorrected and (f) the corrected probe measurements. Only the corrected probe is able to resolve the test object. (g) Cross-sectional profiles across the uncorrected and corrected probes in blue and purple, respectively, averaged over a distance of 60 pm. The maximum intensities of the profiles were normalized to unity so that the diameters can be compared.

A different approach for shaping electron beams could involve an off-axis scheme based on computer-generated holography [23,24] and on the use of binary masks, which can also be used to introduce or negate aberrations. However, phase manipulation is then performed using the first diffraction order, which limits the theoretically accessible intensity to 10% and 40% for binary amplitude masks and phase structures, respectively. All other diffraction orders must then be filtered out from reaching the sample plane, meaning that the installation and operation of such a corrector would require additional optical elements and protocols for STEM imaging, which may not be readily reproducible by the average user [25] and have yet to be demonstrated. Furthermore, the efficiency figures would be diminished by ≈80% due to unavoidable inelastic scattering, reducing the usable intensity and current in the corrected probe. The fabrication of non-binary structures such as blazed gratings is still difficult to implement for truly efficient phase correction [26].

In conclusion, we have shown how the spherical aberration of an electron probe in a STEM can be corrected using a thin membrane of variable thickness, which acts as a foil corrector, providing higher spatial resolution. Our foil corrector can be fabricated in-house and implemented in an existing uncorrected electron microscope without changing the normal operating procedure. Other weakly-scattering materials [27], as well as thinner or fractured (modulo-$2\pi$) designs and spatial filtering at the sample plane, can be used to further reduce the effect of inelastically scattered electrons and to improve STEM image contrast. The concept can be extended to correction of additional aberrations by appropriately designing the thickness profile. Also, while the concept presented here is related to probe correction, a similar approach can in principle also be applied to TEM imaging.

[1]     M. Knoll, E. Ruska, Das Elektronenmikroskop, Zeitschrift Für Phys. 78 (1932) 318–339.
[2]     O. Scherzer, Über einige Fehler von Elektronenlinsen, Zeitschrift Für Phys. 101 (1936) 593–603.
[3]     M. Haider, H. Rose, S. Uhlemann, E. Schwan, B. Kabius, K. Urban, A spherical-aberration-corrected 200kV transmission electron microscope, Ultramicroscopy. 75 (1998) 53–60.
[4]     M. Haider, S. Uhlemann, E. Schwan, H. Rose, B. Kabius, Electron microscopy image enhanced, Nature. 392 (1998) 768–769.
[5]     O.L. Krivanek, N. Dellby, A.R. Lupini, Towards sub-Å electron beams, Ultramicroscopy. 78 (1999) 1–11.
[6]     P.E. Batson, N. Dellby, O.L. Krivanek, Sub-angstrom resolution using aberration corrected electron optics, Nature. 418 (2002) 617–620.
[7]     S.J. Pennycook, The impact of STEM aberration correction on materials science, Ultramicroscopy. (2017).
[8]     H. Boersch, Z. Naturforsch, ber die Kontraste von Atomen in Electronenmikroskop, AA Phys Sci. (1947) 615.
[9]     D. Willasch, High-resolution electron-microscopy with profiled phase plates, Optik (Stuttg). 44 (1975) 17.
[10]    K. Muller, Phase plates for electron-microscopes, Optik (Stuttg). 45 (1976) 73.
[11]    Y. Ito, A.L. Bleloch, L.M. Brown, Nanofabrication of solid- state Fresnel lenses for electron optics, 394 (1998) 49–52.
[12]    R. Shiloh, Y. Lereah, Y. Lilach, A. Arie, Sculpturing the electron wave function using nanoscale phase masks, Ultramicroscopy. 144 (2014) 26–31.
[13]    R. Shiloh, R. Remez, A. Arie, Prospects for electron beam aberration correction using sculpted phase masks, Ultramicroscopy. 163 (2016) 69–74.
[14]    R. Shiloh, A. Arie, 3D shaping of electron beams using amplitude masks, Ultramicroscopy. 177 (2017) 30–35.
[15]    H. Du, A nonlinear filtering algorithm for denoising HR(S)TEM micrographs, Ultramicroscopy. 151

(2015) 62–67.
[16]  W.J. Smith, Modern optical engineering, 3rd ed., McGraw-Hill, 2000.
[17]  C. Boothroyd, A. Kovács, K. Tillmann, FEI Titan G2 60-300 HOLO, J. Large-Scale Res. Facil. JLSRF. 2 (2016) A44.
[18]  V. Ronchi, Forty years of history of a grating interferometer, Appl. Opt. 3 (1964) 437–451.
[19]  J.A. Lin, J.M. Cowley, Calibration of the operating parameters for an HB5 stem instrument, Ultramicroscopy. 19 (1986) 31–42.
[20]  O. Scherzer, The theoretical resolution limit of the electron microscope, J. Appl. Phys. 20 (1949) 20–29.
[21]  T. Malis, S.C. Cheng, R.F. Egerton, EELS log-ratio technique for specimen-thickness measurement in the TEM, J. Electron Microsc. Tech. 8 (1988) 193–200.
[22]  L. Reimer, H. Kohl, Transmission Electron Microscopy, Springer New York, New York, NY, 2008.
[23]  B. Brown, A. Lohmann, Complex spatial filtering with binary masks, Appl. Opt. 5 (1966) 967–969.
[24]  W. Lee, Binary computer-generated holograms, App. Opt. 18 (1979) 3661.
[25]  D. Pohl, S. Schneider, P. Zeiger, J. Rusz, P. Tiemeijer, S. Lazar, et al., Atom size electron vortex beams with selectable orbital angular momentum, Sci. Rep. 7 (2017) 934.
[26]  T. R Harvey, J. S Pierce, A. K Agrawal, P. Ercius, M. Linck, B.J. McMorran, Efficient diffractive phase optics for electrons, New J. Phys. 16 (2014) 93039.
[27]  M. Dries, S. Hettler, T. Schulze, W. Send, E. Müller, R. Schneider, et al., Thin-Film Phase Plates for Transmission Electron Microscopy Fabricated from Metallic Glasses, Microsc. Microanal. 22 (2016) 955–963.
[28]  J.E. Barth, P. Kruit, Addition of different contributions to the charged particle probe size, Optik (Stuttg). 101 (1996) 101–109.
[29]  Z. Yu, P.E. Batson, J. Silcox, Artifacts in aberration-corrected ADF-STEM imaging, Ultramicroscopy. 96 (2003) 275–284.

**Acknowledgments:** This work was supported by the Israel Science Foundation (grant no. 1310/13) and by the German-Israeli Project cooperation (DIP). The research leading to these results has received funding from the European Research Council under the European Union's Seventh Framework Programme (FP7/2007-2013)/ ERC grant agreement number 320832. The authors are grateful to Chris B. Boothroyd, Juri Barthel, Emrah Yucelen and Christian Dwyer for valuable contributions to this work.

## Supplementary Material

1. Spherical phase corrector design
2. Raw data and additional 128 pm measurements
3. Ronchigram evaluation by focal series
4. Detailed corrector fabrication

### 1. Spherical phase corrector design

We previously investigated fractured (modulo-$2\pi$) designs of phase masks [13]. In the present case shown in Fig. S1a, we fabricated continuous, non-fractured refractive foil elements, and restricted ourselves to a maximum thickness of $2t_{2\pi}$. The primary advantages of this approach are the elimination of fabrication difficulties associated with sharp edges (see discussion in [13]), and more relaxed restrictions on the choice of accelerating voltage. Based on Eq. 3, the maximal peak-to-valley height is $\Lambda = 2$, i.e., up to $2\lambda$ of aberrations can be corrected. Although a thicker design would allow stronger correction, it would also result in increased electron scattering and therefore a lower current in the corrected probe. This drawback could be compensated by enlarging the diameter of the corrector or by using a different material [27]. A thinner design would reduce scattering and increase contrast, but would provide less correction. For the present microscope parameters, assuming that the $2t_{2\pi}$ thickness limit can be fully exploited, we use the peak-to-valley formula given previously, $\Lambda = C_s NA^4/16\lambda$, and find that the numerical aperture (NA) for optimal correction is NA=12.4 mrad. The intensity cross-section measurement in Fig. S1b is approximately proportional to the thickness of the membrane, and fits well (dashed red line) to a fourth degree polynomial with $R^2 = 0.986$, as supported by residuals (green line).

Considering only diffraction, chromatic, and spherical aberration, the diameter of the probe is plotted as a function of NA in Fig. S1c. Each diameter is estimated using the 50% current-enclosed criterion, and the different diameters are then combined using a root power sum method [28], as follows. Diffraction broadens the probe for lower values of NA according to the expression $d_{d50} = 0.54\lambda/NA$. For larger values of NA, spherical aberration dominates with a cubic dependence as $d_{s50} = 0.18 C_s NA^3$, while the contribution from chromatic aberration varies linearly as $d_{c50} = 0.34 C_c (\Delta E/E) NA$. As an estimate, we take $C_c = C_s$, and $\Delta E = 0.8 eV$ (as measured in a separate EELS experiment). The combined probe diameter is deduced from the expression $\sqrt{((d_{d50}^4 + d_{s50}^4)^{1.3/4} + d_{I50}^{1.3})^{2/1.3} + d_{c50}^2}$. Broadening due to spatial incoherence is included in the term $d_{I50}$, and is negligible in our experiment as a result of the use of spot size 9 (corresponding to a large demagnification of the source). The results in the design diagram Fig. S1c show an expected optimum (marked "o") for an uncorrected probe at 6.4 mrad with a diameter of 180 pm. The corresponding value for a corrected probe is 12.4 mrad with a diameter of 91 pm (marked "x"). Given that the radius of the corrector $r_{max}$ is fixed, the contribution for the corrector (dashed, triangular-like purple line) is given by calculating the compensated, effective value of $C_s$ for each value of NA: $C_{s,eff}(NA) = C_s + C_{s,\Lambda=2}(NA)$, where $C_{s,\Lambda=2}(NA)$ is given by the peak-to-valley formula, and recalculation of $d_{s50}(C_{s,eff})$. It is important to note that these estimations are guidelines for the design of the corrector, and that rigorous modeling of the probe would require a complete wave-optical numerical simulation. These considerations, as well as possible deviations from the expected values of $C_s$ and $C_c$, an incorrect overall thickness value in the fabrication step, or the use of the corrector at a different accelerating voltage, would only result in a change in the operating numerical aperture. Experimentally, we found an optimum value of NA=7.46 mrad for the uncorrected microscope, and NA=12.1 mrad for the corrected microscope.

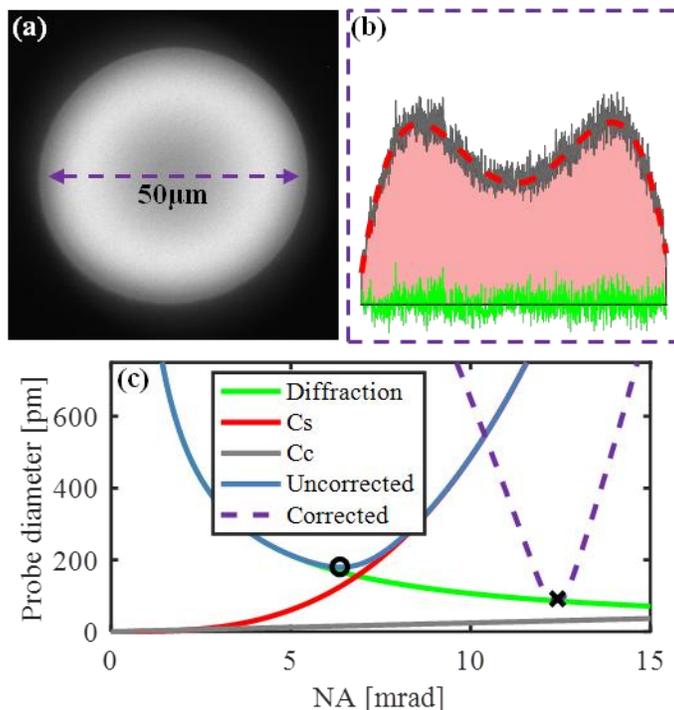

Figure S1: Design and image of a fabricated corrector. (a) Intensity image of the installed corrector recorded in the TEM. Higher intensity corresponds to thinner material. (b) Cross-section of the intensity (black), fit to a fourth degree polynomial (dashed red) and residuals (green). (c) The *geometric* diameter of a probe plotted as a function of the numerical aperture of the probe-forming optics. Diffraction (green) is inversely-proportional to the diameter of the probe, spherical aberration (red) has a cubic dependence, while chromatic aberration (gray) has a linear dependence. The combined effect is calculated as the root power sum (blue line), resulting in an optimum value marked "o" (6.4 mrad / 180 pm). When using the corrector (dashed purple), the optimum value is estimated to be at the point marked "x" (12.4 mrad / 91 pm).

## 2. HAADF-STEM imaging: Raw data and additional 128 pm measurements

HAADF-STEM images of Si {110}, which are shown in Fig. S2, were recorded at 300 keV both with (left column) and without (right column) using the foil corrector. The inner and outer detection angles of the HAADF detector were 50 mrad and 200 mrad, respectively. All of the images are single scans rather than multi-frame averages. A nonlinear filtering algorithm [15] was applied to the raw images both with (Fig. S2a) and without (Fig. S2f) the foil corrector. The same filtering parameters were used in each case to enhance the contrast of the images. Filtered images, which correspond to Fig. 1d and Fig. 1c in the main text, are shown in Fig. S2c and Fig. S2h, respectively. We justify the use of the noise-filtering algorithm in two ways: first, there is no apparent feature in the noise images (Fig. S2e and Fig. S2j), which are subtractions between raw data and filtered data. The filtered data is therefore only random background noise, rather than the addition or subtraction of periodic features. Second, the characteristic spatial frequency, $(136\ pm)^{-1}$, corresponding to the dumbbell separation, can be clearly seen in the Fourier transform power spectra of both the unfiltered (Fig. S2b) and the filtered (Fig. S2d) aberration-corrected images. Without correction, the highest spatial frequencies that are reached in either unfiltered (Fig. S2g) or filtered (Fig. S2i) images are also the same, but are reduced to $(192\ pm)^{-1}$. The resolvable dips between dumbbell peaks in the cross-section profile shown in Fig. 1f in the main text, together with the aforementioned extension in spatial frequency, confirm that the separation of 136 pm dumbbells has been realized using the foil corrector. We note that sub-Angstrom resolution appears to be implied by the presence of a $(96\ pm)^{-1}$ (4-40) spot in Fig. S2d, which, however, is not present in Fig. S2b. A cause could be a nonlinear effect related to the experimental black level, which is discussed comprehensively in [29].

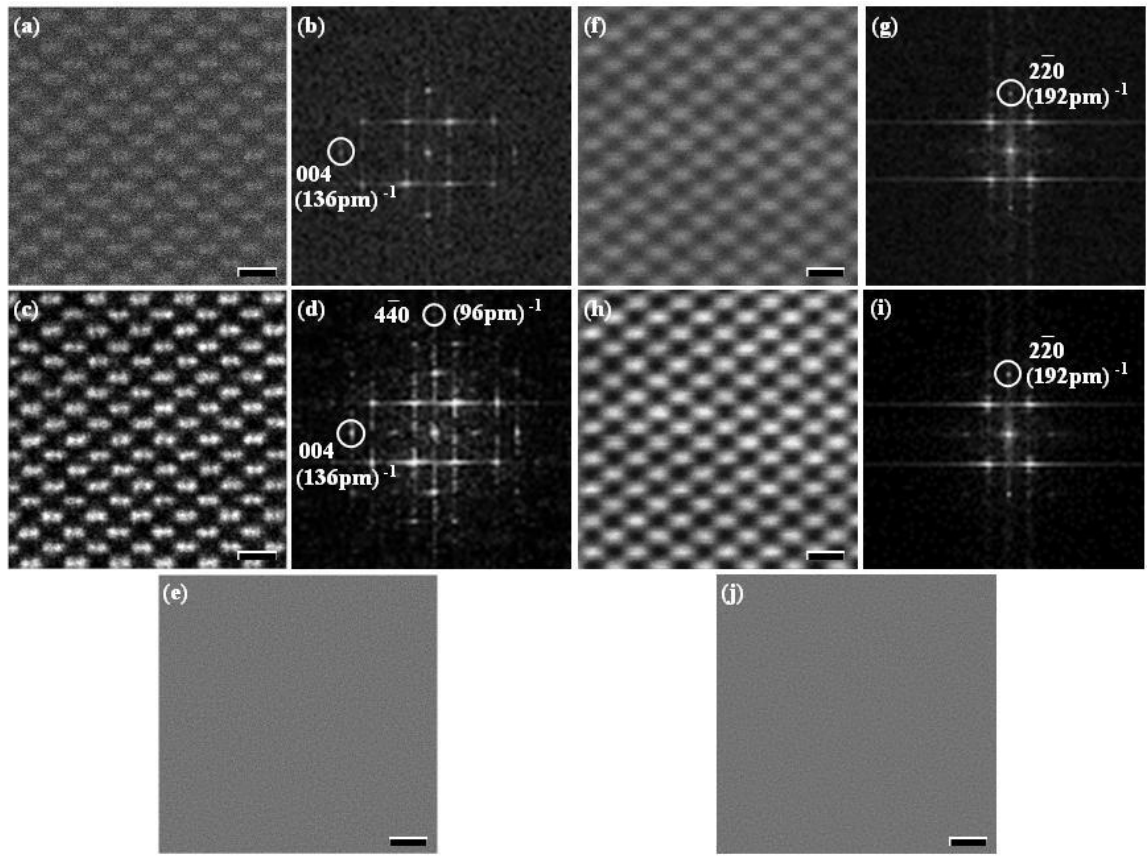

Figure S2: Comparison between raw data (first row) and noise-filtered data (second row) for Si {110} HAADF-STEM images (a, c, f, h) and corresponding Fourier power spectra (b, d, g, i). The difference between unfiltered and filtered images is shown as noise images (e, j) in the third row. The left and right columns correspond to the experimental setup with (a-e) and without (f-j) using the foil corrector. The scale bars in all of the real space images are 500 pm.

A Cu {112} nanocrystalline sample was also measured in HAADF-STEM imaging mode using the foil corrector. Based on the Fourier power spectra shown in Fig. S3b and Fig. S3d, one can notice the presence of a (2-20) spot. In the filtered image shown in Fig. S3c, the 128 pm-separated (2-20) lattice can be clearly resolved in the horizontal arrangement. Although the ability to resolve the periodicity in the raw image is arguable, the filtered image in Fig. S3c, the spectrum in Fig. S3d and the random noise difference image in Fig. S3e demonstrate sub-130 pm resolution.

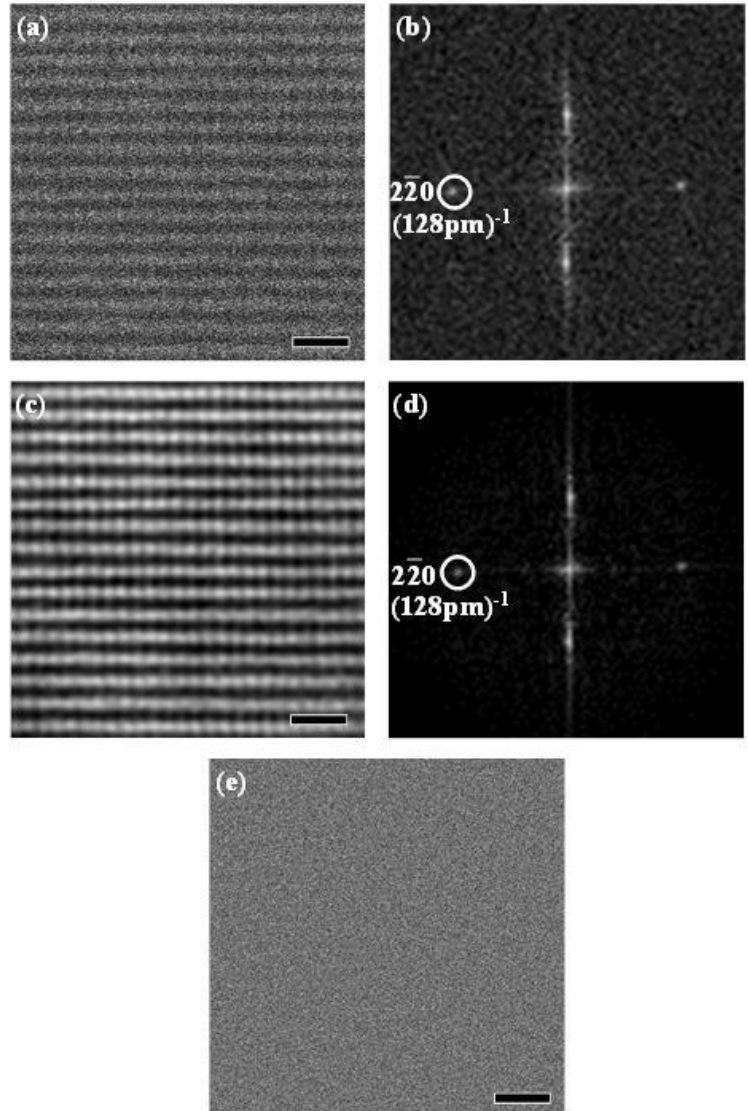

Figure S3: Additional HAADF-STEM images recorded from Cu {112} using the foil corrector, including (a) raw data and (c) a noise-filtered image. The corresponding Fourier power spectra (b, d) both indicate a resolvable 128 pm lattice spacing. The noise image in (e) does not show any apparent feature other than random background noise. The scale bars in all of the real space images are 500 pm.

## 3. Ronchigram evaluation using through-focus series

Fig. S4 shows a focal series of Ronchigrams collected at NA=12.1 mrad, using the foil corrector, from an amorphous C film. At larger defoci (>30 nm), an obvious shadow image is visible in the Ronchigrams, while the image near focus shows infinite magnification and looks smooth and featureless, providing strong evidence for aberration correction up to 12 mrad, which matches our design well. The center is always darker as a result of the thickness modulation.

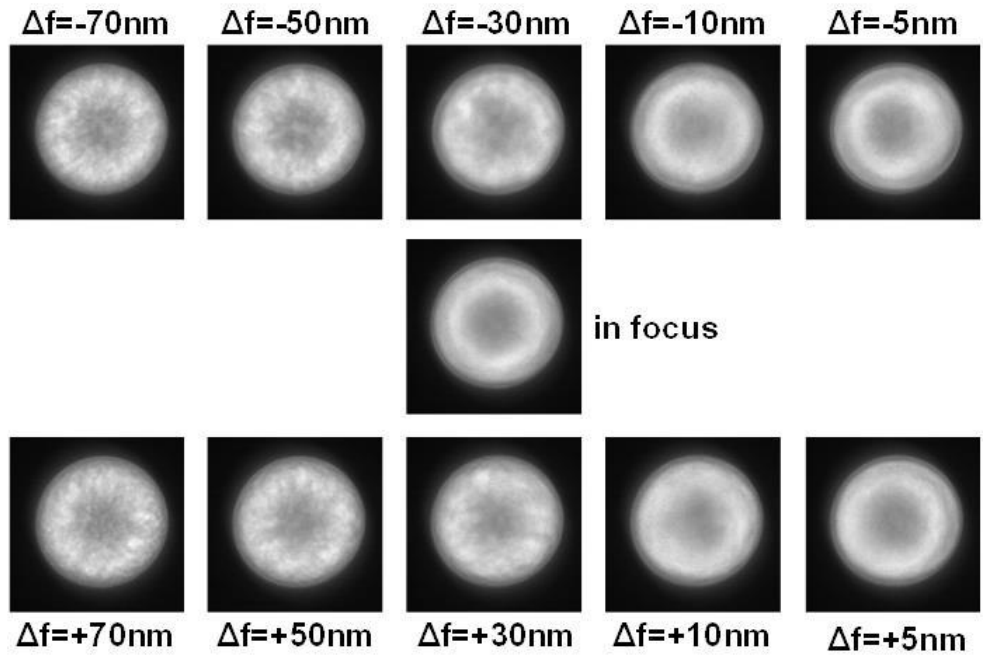

Figure S4: Through-focus series of Ronchigrams recorded from an amorphous C film using the foil corrector. The numerical aperture and cut-off frequency used here is 12.1 mrad.

## 4. Detailed corrector fabrication

Our material of choice for the refractive foil corrector is presently a commercially available 200 nm-thick silicon nitride film, which is based on a 200-$\mu$m-thick Si substrate. We use silicon nitride because it has a good trade-off between scattering, phase-shifting, and mechanical robustness. The window size must be at least twice to three times as large as the corrector diameter, so that electrostatic fields do not develop on the sharp corners of the selectively-etched substrate walls, resulting in high-order aberrations. For a 50-$\mu$m-diameter corrector, a commercial 100 $\mu$m-250 $\mu$m window should then suffice. The membrane is coated on both sides with an adhesive 2 nm of Cr and a thick layer of 150 nm of Au. The purpose of this layer is to reduce excess charge build-up by enhancing conductivity, but, more importantly, to strongly scatter incident electrons, and in principle to act as a strong (opaque) amplitude mask [14], or diaphragm, since beyond the corrector's radius $r_{max}$ the beam retains its high spherical aberration. A round diaphragm, which should be marginally larger than the diameter of the corrector for ease of alignment, is therefore pre-patterned on one side of the membrane using lithography techniques or later removed using FIB milling. On the opposite side, the corrector itself is milled directly through the Au layer, which is easily and completely removed owing to the much higher sputtering rate of Au (compared to silicon nitride). The beam dwell-time during FIB milling translates to depth in an approximately linear manner, although an accurate calibration could be employed, resulting in a surface profile that is defined according to Eq. 3. In our experiments, milling was performed using a Raith IonLine Gallium FIB operated at 35 kV with an 800 pA beam current. As a final step, we coated the membrane with 10 nm of Cr on the flat side for good measure, since bare silicon nitride is insulating.

A commercial-standard 3 mm Si chip includes 9 square windows, each measuring 110 $\mu$m along its edge, meaning that 9 different correctors are available on each chip. In the first stage of our experiments, we noticed a dominant four-fold astigmatism aberration. We hypothesized that the cause of this higher-order aberration was the shape of the window-side of the membrane: that, regardless of the Au layer, the four sharp corners of the Si frame, coupled with the intense electron irradiation, produce electrostatic fields. These fields are thought to symmetrically deflect the probe-forming electrons that pass close to the edges of the window, such that they induce four-fold astigmatism. We corroborated our hypothesis using simulations. A full account of this issue will be published in the future.

An effective solution to this issue is to simply use a commercial Si chip with a larger window that is at least 2-3 times larger than the diameter of the corrector, as previously stated. However, since we installed our corrector on the Condenser 2 aperture holder - a day-long process, due to time constraints, we preferred having a similar 9-window chip but with larger windows, thus experimenting with nine correctors per chip rather than one. To this end, we used a photolithography process to create our own chips with 9 window membranes, but with a larger window area. We began with a 250 $\mu$m Si wafer coated with 200 nm low-pressure chemical vapor-deposited silicon nitride. Using photolithography, we defined apertures on one ("flat") side, coated with Cr and Au and performed lift-off, then defined the windows on the other ("window") side. The wafer was then initially etched through ≈230 $\mu$m of its thickness using a deep-reactive ion-etching machine. The process was completed by a couple of hours of KOH etching. The membranes were then coated from the window side with a Cr-Au layer, and the corrector was fabricated using FIB milling, as explained above. We produced both 50 $\mu$m-diameter and 100 $\mu$m-diameter correctors in this way; the differences between a commercial membrane (≈110 $\mu$m edge, Fig. S5a) and our membrane (≈230 $\mu$m, Fig. S5b) are shown in the form of optical microscope images. In Fig. S5b, one can notice the faint circumference of a 120 $\mu$m aperture fabricated on the flat side of this specific membrane in the photolithography step.

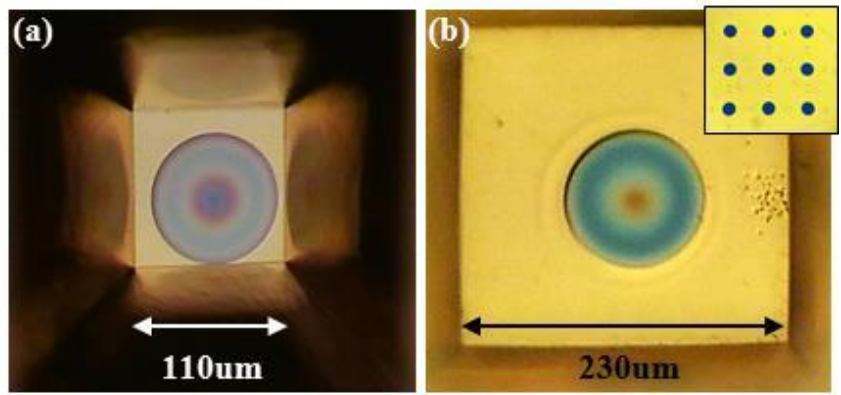

Figure S5: Optical microscope images of the window side of one window of (a) a commercial membrane with a 110 $\mu$m edge, and (b) our in-house-fabricated membranes with a 230 $\mu$m edge. Inset: flat side showing 9 open apertures of bare silicon nitride (blue) in the Au layer (yellow). Both are 200 nm silicon nitride, coated with Cr and Au from the window side, and show a milled 100 $\mu$m corrector. In (b), there is faint evidence of the 120 $\mu$m-diameter aperture on the flat side's Au layer. Varying colors result from different illumination conditions. The contrast and brightness have been changed for visibility. In (a), we did not have a Au layer on the flat side.